# Solar Ring Mission: Building a Panorama of the Sun and Inner-heliosphere

Wang, Yuming[1,2,*](ymwang@ustc.edu.cn); Bai, Xianyong[3](xybai@bao.ac.cn); Chen, Changyong[4](chency@microsate.com); Chen, Linjie[3](ljchen@nao.cas.cn); Cheng, Xin[5](xincheng@nju.edu.cn); Deng, Lei[4](deng_leo@163.com); Deng, Linhua[6](lhdeng@ynao.ac.cn); Deng, Yuanyong[3](dyy@nao.cas.cn); Feng, Li[7](lfeng@pmo.ac.cn); Gou, Tingyu[2](tygou@ustc.edu.cn); Guo, Jingnan[1,2](jnguo@ustc.edu.cn); Guo, Yang[5](guoyang@nju.edu.cn); Hao, Xinjun[1,2](xjhao@ustc.edu.cn); He, Jiansen[8](jshept@pku.edu.cn); Hou, Junfeng[3](jfhou@bao.ac.cn); Huang, Jiangjiang[4](huangjj@microsate.com); Huang, Zhenghua[9](z.huang@sdu.edu.cn); Ji, Haisheng[7](jihs@pmo.ac.cn); Jiang, Chaowei[10](chaowei@hit.edu.cn); Jiang, Jie[11](jiejiang@buaa.edu.cn); Jin, Chunlan[3](cljin@nao.cas.cn); Li, Xiaolei[1,2](lxllxl@ustc.edu.cn); Li, Yiren[1,2](lyr@ustc.edu.cn); Liu, Jiajia[12](J.Liu@qub.ac.uk); Liu, Kai[1,2](kailiu@ustc.edu.cn); Liu, Liu[4](liul@microsate.com); Liu, Rui[2](rliu@ustc.edu.cn); Liu, Rui[4](liur@microsate.com); Qiu, Chengbo[4](qiucb@microsate.com); Shen, Chenglong[1,2](clshen@ustc.edu.cn); Shen, Fang[13](fshen@spacewether.ac.cn); Shen, Yuandeng[6](ydshen@ynao.ac.cn); Shi, Xiangjun[6](shixiangjun@ynao.ac.cn); Su, Jiangtao[3](sjt@bao.ac.cn); Su, Yang[7](yang.su@pmo.ac.cn); Su, Yingna[7](ynsu@pmo.ac.cn); Sun, Mingzhe[9](sunmingzhe@sdu.edu.cn); Tan, Baolin[3](bltan@nao.cas.cn); Tian, Hui[8](huitian@pku.edu.cn); Wang, Yamin[4](wangym@microsate.com); Xia, Lidong[9](xld@sdu.edu.cn); Xie, Jinglan[6](xiejinglan@ynao.ac.cn); Xiong, Ming[13](mxiong@swl.ac.cn); Xu, Mengjiao[1,2](xmj0517@ustc.edu.cn); Yan, Xiaoli[6](yanxl@ynao.ac.cn); Yan, Yihua[3](yyh@bao.ac.cn); Yang, Shangbin[3](yangshb@nao.cas.cn); Yang, Shuhong[3](shuhongyang@nao.cas.cn); Zhang, Shenyi[13](zsy@nssc.ac.cn); Zhang, Quanhao[1,2](zhangqh@ustc.edu.cn); Zhang, Yonghe[4](zhangyh@microsate.com); Zhao, Jinsong[7](js_zhao@pmo.ac.cn); Zhou, Guiping[3](gpzhou@nao.cas.cn); Zou, Hong[8](hongzou@pku.edu.cn)

[1] Deep Space Exploration Laboratory/School of Earth and Space Sciences, University of Science and Technology of China, Hefei 230026, China
[2] CAS Center for Excellence in Comparative Planetology/CAS Key Laboratory of Geospace Environment/Mengcheng National Geophysical Observatory, University of Science and Technology of China, Hefei 230026, China
[3] National Astronomical Observatories, Chinese Academy of Sciences, Beijing, China
[4] Innovation Academy for Microsatellites, Chinese Academy of Science, Shanghai, China
[5] Nanjing University, Nanjing, China
[6] Yunnan Observatories, Chinese Academy of Sciences, Kunming, China
[7] Purple Mountain Observatory, Chinese Academy of Sciences, Nanjing, China
[8] Peking University, Beijing, China
[9] Shandong University, Weihai, China
[10] Harbin Institute of Technology, Shenzhen, China
[11] BeiHang University, Beijing, China
[12] Queen's University Belfast, Belfast, UK
[13] National Space Science Center, Chinese Academy of Sciences, Beijing, China
* Corresponding Author (ymwang@ustc.edu.cn)

## Abstract

Solar Ring (SOR) is a proposed space science mission to monitor and study the Sun and inner heliosphere from a full 360° perspective in the ecliptic plane. It will deploy three 120°-separated spacecraft on the 1-AU orbit. The first spacecraft, S1, locates 30° upstream of the Earth, the second, S2, 90° downstream, and the third, S3, completes the configuration. This design with necessary science instruments, e.g., the Doppler-velocity and vector magnetic field imager, wide-angle coronagraph, and in-situ instruments, will allow us to establish many unprecedented capabilities: (1) provide simultaneous Doppler-velocity observations of the whole solar surface to understand the deep interior, (2) provide vector magnetograms of the whole photosphere — the inner boundary of the solar atmosphere and heliosphere, (3) provide the information of the whole lifetime evolution of solar featured structures, and (4) provide the whole view of solar transients and space weather in the inner heliosphere. With these capabilities, Solar Ring mission aims to address outstanding questions about the origin of solar cycle, the origin of solar eruptions and the origin of extreme space weather events. The successful accomplishment of the mission will construct a panorama of the Sun and inner-heliosphere, and therefore advance our understanding of the star and the space environment that holds our life.

1. Background and Scientific Rationale

As the nearest star from the Earth and the major energy source of the solar system, the Sun has been the eternal pursuit of humans to be understood. Significant advances in the theory and observation of the Sun have been seen in the past decades, especially thanks to the rapid development of solar and heliophysics missions. More than 70 missions have been successfully launched in the past 50 years, among which about 10 are still actively observing the Sun. These missions include but are not limited to SOlar and Heliospheric Observatory (SOHO, Domingo et al. 1995), PROBA 2 (Berghmans et al. 2006), Hinode (Kosugi et al. 2007), Solar TErrestrial RElations Observatory (STEREO, Kaiser et al. 2008), Solar Dynamics Observatory (SDO, Pesnell et al. 2012), Interface Region Imaging Spectrograph (IRIS, De Pontieu et al. 2014), Solar Orbiter (Müller et al. 2020) and Parker Solar Probe (PSP, Fox et al. 2016), and Chinese Hα Solar Explorer (CHASE, Li et al. 2022). These missions cover a wide range of detections of the Sun and the interplanetary space by utilizing multi-wavelength and multi-layer imaging, and in-situ observations of the magnetic field and plasma parameters. Via numerous studies of the solar quiet and active regions, the solar activity cycle, solar wind and solar eruptions, countless important advances have been made in understanding solar internal structures, the coronal heating problem, the origin of the solar wind, the generation mechanism and propagation process of solar eruptive events, and the origin and prediction of severe space weather events.

At the time scale of climate from tens to thousands of years, the long-term evolution of the terrestrial space environment is controlled by the long-term solar activity cycles. It is widely believed that the origin of solar activity cycles is closely related to the solar dynamo (Karak et al. 2014), which is the MHD process as a result of the non-linear interaction between turbulences and the magnetic field in the solar convection zone. A detailed study of the origin and evolution of solar activity cycles relies on the measurement of the solar differential rotation and the inversion of the status of the magnetic structures and plasma fluid of the convection zone. Therefore, it has been one of the main means to understand solar activity cycles via observing the Doppler velocity and magnetic field of the solar surface. For example, the Michelson Doppler Imager (MDI, Scherrer et al. 1995) onboard SOHO which was launched in 1995, for the first time provided us with long-term observations of the Earth-facing side of the Sun. From 2012, the Helioseismic and Magnetic Imager (HMI, Schou et al. 2012) onboard SDO started to provide the vector magnetic field and Doppler velocity observation of the Earth-facing side of the Sun with high spatial and temporal resolutions. These observations have provided a huge amount of important data to understand solar activity cycles (Yang & Zhang 2014, Xie et al. 2017, Jiang & Cao 2018), utilizing which people were able to obtain the profiles of the solar internal differential rotation with relatively high resolution and the structural characteristics of the tachocline at the bottom of the convection zone that is widely believed to be generated by the toroidal field. However, the distribution of the solar internal meridional circulation is still under intense debate, mainly due to the limit of single-point observations. Thus, multi-vantage observations are needed to effectively reduce systematic errors in the data, improve the signal-to-noise ratio (SNR) in the detections of the convection zone and therefore advance our understanding of the solar cycles.

At the time scale of weather from minutes to days, the terrestrial space environment is also significantly affected by solar eruptions. Understanding the origin of solar eruptions is one of the keys to understanding and predicting short-term changes in interplanetary and planetary space. A number of breakthroughs have been made in this aspect in recent years. Hinode, launched in 2006, was the first space-borne mission to measure the high-resolution vector magnetic field of the photosphere, making the 3D non-linear magnetic field reconstruction and the quantitive study of the magnetic energy variation during solar eruptions possible (Kosugi et al. 2007). In the same year, the US National Aeronautics and Space Administration (NASA) launched STEREO, initiating the first-ever stereoscopic observations of the Sun (Kaiser et al. 2008) and allowing the reconstruction of the three-dimensional structures of various coronal features (Veronig et al. 2008). SDO, launched in 2010, provides observations of the Sun with higher spatial and temporal resolutions in more ultraviolet passbands (Pesnell et al. 2012). Using these observations, it was found that the precursor structures of solar eruptions are magnetic flux ropes with temperatures up to tens of million degrees (Zhang et al. 2012, Liu 2020). Its high-resolution imaging has also revealed the two wavefronts of coronal mass ejections (CME), resolving the long-standing dispute over the nature of coronal extreme ultraviolet waves (Patsourakos & Vourlidas 2012). In addition, SDO provides uninterrupted observations of the solar photospheric vector magnetic field (Schou et al. 2012), allowing people to perform 3D magnetic field extrapolations to quantify the evolution of the magnetic field energy and topology during solar eruptions (Sun et al. 2012). Numerous models of solar eruptions have since been proposed and tested in observations and numerical simulations, among which some have successfully reproduced many observational features of the eruptions (Kliem et al. 2013, Guo et al. 2019, Zhong et al. 2021).

On the other hand, extreme space weather events may be triggered by the complex interactions between violent solar eruptions and the Earth's magnetosphere (e.g., Wang et al. 2003, Shen et al. 2017, Temmer 2021). They may have disastrous impacts on the Earth's space environment and endanger the safety of human infrastructures and activities including aviation, spaceflight, communication, and power grids. Space weather events include two main effects: solar energetic particles (SEPs) and geomagnetic storms, which depends on CMEs properties. Therefore, knowing the acceleration and propagation characteristics of CMEs is a key to understanding and predicting extreme space weather events. For a long time, the study of CMEs suffers from the line-of-sight projection effect and their evolution in the interplanetary stays unclear, mainly because only single-point observations are available. Since 2006, the stereoscopic observations of CMEs from the Sun to the interplanetary space became possible after the launch of STEREO, leading to many advances in the understanding of CMEs including the evolution of their speed (Bein et al. 2011), the evolution of their propagation (Mostl et al. 2015), their reformation (Riley and Crooker 2004), their interaction with each other (Shen et al. 2012, Lugaz et al. 2017) and their geomagnetic effects (Scolini et al. 2020). Several prediction models have been developed as a result (Vršnak et al. 2013, Riley et a. 2018). Meanwhile, the use of multi-vantage observations has led to a greater awareness of solar energetic particle events, particularly about their diffusion and propagation in interplanetary space. A solar energetic particle event with an extension of more than 180° in the interplanetary space was observed for the first time (Rodríguez-García et al. 2021).

Despite many achievements having been made, new issues are continuously emerging. Firstly, continuous observations of the full-surface solar magnetic field and velocity field are needed to uninterruptedly trace sunspots, prominences and active regions, and thus to explore the evolution of solar activity cycles. Current understandings of the physics of these phenomena are limited because single-point observations have a continuous tracking capability of fewer than 13 days, well below the lifetime of these phenomena. Secondly, real-time global magnetic field observations are key to understanding the global image of the accumulation and release of magnetic energy in the solar atmosphere. However, current techniques only allow mosaic observations via joining individual magnetic maps obtained across 27 days, which are usually inaccurate and seriously affect our understanding of the generating mechanisms of solar internal magnetic fields and eruptive events. These synoptic observations also fail in providing accurate boundary conditions for data-driven global MHD simulations which rely on real-time 3D coronal magnetic field parameters. Thirdly, the low accuracy of transverse magnetic field and 180° ambiguity in single-point measurements greatly limit the accuracy of coronal magnetic field extrapolations, which can be effectively mitigated in multi-vantage magnetic field observations. Fourth, there are significant line-of-sight integral and projection effects of the very thin plasma in interplanetary space, which can again be greatly relieved in multi-vantage observations. Fifth, the large extension of interplanetary shocks allows accelerated particles to be distributed in a wide longitudinal range, thus extensively influencing the interplanetary space. The deployment of several large-angle separated satellites on the ecliptic plane will enable the analysis of the longitudinal distribution of solar disturbance events and further provide more effective support for space weather forecasting.

Thus, we propose the Solar Ring (SOR) mission, which is designed to deploy three (S1, S2 and S3) 120°-separated spacecraft (Figure 1) from 30° upstream of the Earth on the 1 AU orbit around the Sun in the ecliptic plane, to enable multi-vantage panoramic observations and studies of the Sun and the interplanetary space. This is an optimized version of our previously proposed SOR mission (Wang et al. 2020, 2021), in which 6 spacecraft (3 groups of two 30°-separated spacecraft) were suggested. Compared with the previous design, the deployment of three spacecraft will achieve almost the same scientific objectives but with much lower cost.

The design of SOR is not like STEREO, which has two spacecraft continuously drifting away from the Earth. No full Sun view can be steadily and continuously obtained by STEREO. Particularly, STEREO has no observation of the photospheric magnetic field, which is very important for us to advance our understanding of solar cycles and solar eruptions. This design is also better than that of L5 and/or L4 mission if not considering the cost. The most notable advantage is that SOR can have a panoramic view of the Sun and inner heliosphere that a L5/L4 mission cannot achieve. It will be a great leap for many solar physics issues as described in Sec.2 and Sec.5. Another advantage is that S1 is only 30° upstream of the Earth, which allows the space weather forecast more accurately than that from larger separation angles, such as the around 60° of L5 mission (Chi et al. 2022)

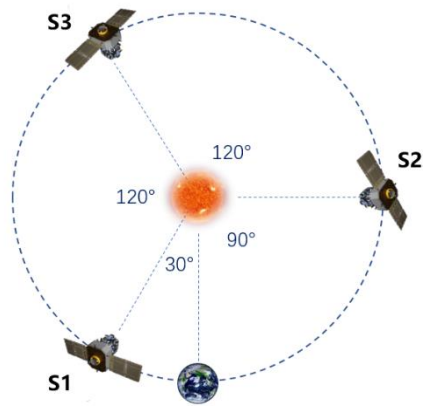

Figure 1. Configuration of SOR mission.

## 2. Scientific Questions and Mission Objectives
### 2.1 Scientific Questions

**In the Time Scale of Climate, How and Why Do the Solar Cycles Evolve?** The solar system is the only known celestial system that has a habitable planet, of which the habitability is by large determined by the activity level of the Sun. Though it is widely agreed that human activities have been the main reason for global warming in the past decades, the long-term evolution of Earth's climate is determined by the long-term variation of solar activities (Pustilnik and Din 2004). The lack of knowledge of the long-term variation of solar activities limits our understanding of the past, present and future of the Earth, given that recorded sunspot observation only dates back to 400 years ago and the total solar irradiance derived from the isotopic abundances of cosmic rays covers merely 11,500 years (Vieira et al. 2011), which are extremely short compared to the history of the Earth and solar system. Therefore, it is a fundamental and key scientific question that how we may proceed in understanding the physics and rules of the long-term evolution of solar activities via modern observations, to infer the past and further predict the future of the Sun. To answer this essential scientific question, we need to have a deeper understanding of the mechanisms and evolution of solar activities, through intensive studies of solar internal structures, their dynamics, the generation and emergence of magnetic fluxes, and the distribution and temporal evolution of the solar global magnetic field.

**In the Time Scale of Weather, How and Why Do the Solar Eruptions Occur and Influence the Space Weather?** Solar eruptions are the most violent release of material and energy in the solar system. A typical solar flare or CME can inject $10^9$ tons of magnetized plasma, $10^{25}$ joules of energy and a large number of energetic particles (Benz 2017) into the interplanetary space in a short time. They occur about once every two days during the solar minima and on average 4 to 5 times a day during solar maxima. These violent outbursts from the Sun significantly disturb interplanetary space, leading to potentially catastrophic space weather events. It normally takes around 8 minutes for the electromagnetic radiation, around 1 hour for energetic particles and one to several days for the plasma ejecta to reach the Earth from the Sun. Thus, how to predict solar eruptive activities and extreme space weather events is an important scientific proposition to ensure the safe operation of human high-tech systems and deep space exploration missions. However, it is still impossible for humans to make long-term or even short-term predictions of space weather as accurate as the weather forecast. While the key lies in the understanding of the physics and mechanisms of the accumulation and release of magnetic energy during solar eruptions, and the coupling processes of plasma and magnetic field in different space-time scales. To answer this key scientific question, we need to obtain information about the whole Sun from a panoramic view to advance our understanding of how solar eruptions couple with the global coronal and interplanetary magnetic field (IMF), how the energy is built up, stored and released during eruptive events, and how energetic particles are generated and accelerated.

**In Terms of the Spatial Distribution, How Do the Solar Wind Disturbances and Energetic Particles Propagate in the Heliosphere?** The transport and diffusion of waves or particles during solar eruptive events in interplanetary space are among the most spectacular phenomena in the solar system. A large number of observations suggest that rapid ejecta, along with shock waves driven and energetic particles accelerated by them, can have longitudinal and latitudinal extensions up to more than 180°, thus further affecting more than one planet in the solar system (Guo et al. 2018; Prangé et al. 2004). In the vast expanse of space, how plasma contained in solar eruptive events are diffused and transported determines properties, especially intensities, of disturbances in interplanetary space, and thus affects the implementation of human deep space exploration missions. In the case of a Mars mission, it usually takes 7 to 8 months for a spacecraft to arrive on Mars via the Hohmann transfer orbit. The spacecraft has to wait for another 15 months before it travels for 7 to 8 months again to arrive on Earth via the Hohmann transfer orbit. Among its 2-3 years of travel in deep space and Mars' surface, predicting energetic particle radiation and protecting the spacecraft from it will be a huge challenge, especially for manned exploration missions. Note that all planets and most celestial bodies in the solar system are located near the ecliptic plane, it is then an especially important scientific proposition to conduct 360° panoramic observations of the Sun and the interplanetary space near the ecliptic plane, to fully understand the diffusion laws and transport mechanisms of solar eruptive events in both longitudinal and latitudinal directions, and to guarantee the success of human's future deep space exploration missions. However,

because of the vastness of interplanetary space, the distribution of violent disturbances caused by solar eruptions in interplanetary space is still not fully understood. Many questions remain unanswered, including how solar wind disturbances evolve in interplanetary space, how SEPs travel through interplanetary space, and what are the main parameters and patterns of the interplanetary disturbances that mostly affect the space environment?

## 2.2 Mission Objectives

The core scientific goal of SOR is to achieve the first panoramic observations of the Sun and the inner heliosphere for updating our knowledge of the star and the space environment that hold our life. It aims to tackle three fundamental "origin" problems in solar and space physics:

**Origin of solar cycles.** The first large-angle multi-vantage observations of physical parameters such as the Doppler velocity field and vector magnetic field of the solar surface will be carried out, to study solar internal structures, its internal transport processes, and the evolution of solar surface features, and to further understand and predict solar cycles. These will also deepen our understanding of the evolution of Sun-like stars and the habitability of their planets.

**Origin of solar eruptions.** SOR will study how magnetic energy is accumulated, stored and released, how large-scale eruptive structures are formed and developed, and how the coronal magnetic field evolves during eruptive events, through ultraviolet imaging and accurate measurements of the photospheric vector magnetic field from more than one perspective. These will advance our understanding of solar eruptive events and help us make more accurate predictions.

**Origin of extreme space weather events.** Global observations of the solar wind in the middle and low latitudes of the inner heliosphere will be realized for the first time through wide-angle multi-vantage imaging of the corona and the interplanetary space with a large field of view. Combined with multi-vantage in-situ measurements, the formation, evolution, longitudinal distribution, and space weather effects of solar energetic phenomena will be studied, to enrich our understanding and enhance our prediction abilities of extreme space weather events.

## 3. Scientific Requirements and Instruments
## 3.1 Scientific Requirements

To tackle the aforementioned three fundamental "origin" problems, the Solar Ring mission will carry 10 different scientific payloads, including 6 remote sensing instruments: the Spectral Imager for Magnetic field and helioSeimsology (SIMS), Multi-band Imager for Extreme ultraviolet emissions (MIE), High-energy Emission Monitor for flares (HEM), Integral Spectrograph for Extreme ultraviolet Emissions (ISEE), Wide-Angle Coronagraph (WAC), low-frequency radio investigator (WAVES), and 4 in-situ measurement instruments: magnetometer (MAG), Solar wind Plasma Analyzer (SPA), Medium-energy Particle Detector (MiPD), High-energy Particle Detector (HiPD). The corresponding scientific motivation and instrumental requirements of SOR for each of the "origin" problems are briefly discussed below.

**Origin of the solar cycles.** First, we need to study the internal structure and dynamic process of the sun for which observations of helioseismology are essential. To reduce the problem of spectral leakage in global helioseismology and to effectively remove systematic errors such as the center-edge differences and projection effects, it is necessary to measure the full-disk Doppler velocity field in 360 degrees to invert the internal structure and dynamic process. Second, to study the generation, emergence and evolution of global magnetic flux, it is necessary to continuously track the magnetic field and flow field of magnetic structures at different scales such as sunspots. This also requires 360° observation of full-disk Doppler velocities and vector magnetic field to be realized by SIMS at all three spacecraft.

**Origin of solar eruptions.** First, we need to study coronal energy accumulation and release processes and how they are coupled with the global coronal and IMF. It is thus necessary to carry out data-driven global MHD simulations of solar eruptions by reconstructing the three-dimensional coronal magnetic field structure which will be compared with extreme ultraviolet (EUV) and white-light images. Second, to understand the energization mechanism of SEPs, we need to analyze the three-dimensional structure and the evolution of the energy-releasing sites using EUV, X-ray and radio emissions. Above scientific requirements need to be met using synergistic observations at multi-vantage spacecraft using SIMS, WAC, MIE, HEM, ISEE, WAVES, MiPD and HiPD.

**Origin of extreme space weather events.** First, it is important to study the propagation and evolution of solar wind disturbances from different viewing angles using coronagraph imagers together with radio and in-situ measurements. Second, to study the transport of SEPs in the interplanetary space, we need to use multi-view (and thus three-dimensional) observations of CMEs, shocks, CIRs together with in-situ spectral and flux observations of SEPs. Third, to study the near-Earth space environment and forecast magnetic storms, we need to rely on the upstream remote-sensing and in-situ observations of the Sun and IMF. To achieve the above scientific requests, synergistic observations from multi-viewpoints by SIMS, WAC, MIE, HEM, ISEE, WAVES, MAG, SPA, MiPD and HiPD are needed.

Most of the ten payloads will be carried by all three spacecraft. However, due to satellite resource, telemetry and cost constraints, MIE is only onboard the near-Earth spacecraft S1 (see Figure 1); HEM is only at S1 and S2, while S3 is compromised with ISEE consuming much less resource. ISEE is also mounted on S1 for comparative studies and cross-calibration purposes. The remaining seven payloads are mounted on all three spacecraft with different resolutions and/or cadences. The detailed requirements of payloads defined by the key scientific goals, questions and analyzing methods are tabulated in Table A1 in the Appendix.

### 3.2 Key Requirements of the Instruments

Below, we discuss in better detail of some key scientific requirements of these instruments.

**Spectral Imager for Magnetic field and helioSeimsology (SIMS).** To minimize systematic errors such as spectral leakage, center-edge differences and projection effects, it requires multi-view 360° observations with a temporal resolution of about 1 minute. Moreover, the detection of deeper structures such as the convection zone and global dynamics requires relatively long-lasting continuous observations (36 to 72 days) with a spatial resolution of up to about 5000 km. For photospheric vector magnetic fields, an observation wavelength of 5324 Å is recommended. Since the time scale of the accelerated phase of solar eruptions is typically 10-20 minutes, we adopt a resolution of 15 minutes to measure the vector magnetic field on S1. While for S2 and S3 which are further away, the resolution is compromised to be 60 minutes and 120 minutes, respectively. To resolve small-scale kinematic magnetic features that trigger solar eruptions, the spatial resolution is recommended to be better than 2 arcseconds. To mitigate the issues of low accuracy in the transverse magnetic fields and 180-degree uncertainty, Ji et al. (2019) and Zhou et al. (2021) demonstrated that both problems can be effectively resolved when the magnetic fields are measured simultaneously in two directions separated by about 30°.

**Multi-band Imager for Extreme ultraviolet emissions (MIE).** Reconstructing the three-dimensional structure of solar coronal features also requires two views for EUV imaging with their angular separation between 20° and 60°. This will be achieved by synergistically using existing measurements on spacecraft near the Earth such as SDO and ASO-S. The SOR S1 will be placed 30° upstream of the Earth allowing joint observations of the vector magnetic field and reconstruction of the 3D structure of the coronal features. To resolve the structure of the precursors of solar eruptions the EUV imager on S1 requires a spatial resolution of about 3 arcseconds that can be realized by a 2K × 2K CCD. The rising and accelerating phases of solar eruptions last about a few minutes which requires a temporal resolution of at least 10 seconds. Considering the data limit of telemetry, two modes will be adopted for downloading data: quiet mode and flare mode. The former has a time resolution of 1 minute and the latter has a time resolution of 2 seconds. In addition, to capture the global perturbations and early evolution of CMEs, a large observational field of view is required, and 64 × 64 arcmin2 is recommended. Considering that the solar atmosphere covers a temperature range from 10,000 K (chromosphere) to 20 million K (flares in the corona), we will use three bands, 304, 174, and 130 Å, to image the chromosphere, corona, and flare plasma simultaneously.

**High-energy Emission Monitor for flares (HEM).** HEM needs to have both energy spectrum detecting capability and imaging capability, to monitor solar flares and their three-dimensional structures and to study flare radiation and its anisotropy as well as particle acceleration and propagation processes. Considering the limitations of data transmission, spacecraft weight and size, the HEM of S1 can obtain the energy spectrum, X-ray imaging and image spectroscopy from soft X-rays to hard X-rays, with a field of view covering 1.5 times of the solar radius. The proposed spatial resolution is better than 8 arcseconds and temporal resolution is better than 2 seconds. For better observations of higher energy electrons and ions, HEM also can obtain gamma-ray energy spectra (up to 10 MeV). To resolve the Fe-Ni emission line at 6-8 keV, a spectral resolution better than 1 keV@ 6 keV is recommended. For S2, due to further resource constraints, we use a high-energy spectrometer for basic X-ray imaging and gamma-ray spectral observations with a spatial resolution better than 30 arcsec.

**Integral Spectrograph for Extreme ultraviolet Emissions (ISEE).** To explore the detection methods of stellar CMEs, ISEE is considered for S3 and the near-Earth spacecraft S1. Coronal mass ejections typically exhibit blue-shifted features in the ultraviolet spectrum (Tian et al. 2012), and the proposed observing band range is 180-270 Å with a spectral resolution better than 500. S1 will pair EUV imaging with a high-resolution integral spectrometer (ISEE) for the first time to study in detail the origin and evolution of CMEs in the integral spectrum, providing a unique reference for stellar CME detections. The S3/ISEE can detect the process of the backside (as seen from Earth) flare generation and CME eruption, measure the apparent velocity of backside CMEs, constrain the backside eruptions in global low coronal models. One can also apply the imaging-spectroscopy correlation established based on S1 ISEE and MIE observations to S3 observations to infer the EUV structure and evolution characteristics seen by S3.

**Wide-Angle Coronagraph (WAC).** We have demonstrated that the identification and reconstruction of the solar wind disturbance structure are optimized when multi-view observers are separated by about 120° as designed for SOR. Meanwhile, the continuous solar wind observations by S1 30° upstream of the Earth can be used for forecasting geomagnetic storms about 50 hours in advance with high accuracy. To monitor the propagation of solar wind disturbance structures such as CMEs, shocks and small-scale blobs, we need a large field of view white light coronagraph observations by WAC to obtain information on the geometry and propagation dynamics of solar wind structures. Since the main evolution of the solar wind structure occurs in the region close to the Sun, WAC has a field of view of 5-45 Rs and a stray light suppression level of 10-12 in the outer field of view.

**Low-frequency radio investigator (WAVES).** To resolve the electron beams that generate the interplanetary radio bursts accompanying solar eruptions, WAVES will cover the range of 10 kHz - 50 MHz, a frequency band with a radiation source region covering the solar mid-corona to near-Earth space. It has a temporal resolution of 1 s and a frequency resolution of about 0.3% of the central frequency, allowing full resolution of the time scale and spatial scale of the eruptions in this band. This configuration is sufficient to track the details of the propagation and evolution of CMEs and SEPs in interplanetary and near-Earth space.

**Magnetometer (MAG) & Solar wind Plasma Analyzer (SPA).** To analyze the in-situ characteristics of the solar wind structure, we need both magnetometer (MAG) and solar wind plasma analyzer (SPA). MAG has a detection range of ±2000 nT with a resolution better than 0.01 nT and it will provide a variety of sampling rates: 1, 32, 128 Hz. SPA detects particles in the energy range of 100 eV – 30 keV with a mass range of 1-70 amu. The time resolution is 32 s.

**Medium and High-energy Particle Detectors (MiPD & HiPD).** To analyze the generation and propagation of SEPs in interplanetary space, we need information on energy spectra and anisotropy of intermediate and high energy electrons and protons, as well as high energy heavy ions. We also need to measure radiation dose rate as well as the LET (Linear energy transfer) histogram to forecast the radiation level for 1 AU orbit. MiPD will cover an energy range of 0.04-0.6 MeV for medium-energy electrons. HiPD will cover 0.1-12 MeV for high-energy electrons, 0.05-300 MeV for medium- and high-energy protons, and 10-100 MeV for high-energy heavy ions.

The 10 instruments are proposed to meet the scientific requirements of the SOR mission. The detailed information, including the design, technical specifications, technology readiness level (TRL) and so on, has been summarized in Table A2 in the Appendix. Although China has the ability to design and construct all the instruments regardless of the TRL, we welcome any international partners to involve in this mission. All the scientific payloads will be subject to global public bidding.

## 4. Mission Profile and Design
### 4.1 Launch and Orbit

The SOR mission intends to simultaneously launch three spacecraft into the Earth-moon transfer trajectory in a way of "three spacecraft with one rocket" using an LM-3B launch vehicle, from the Xichang satellite launch center. With the help of the moon's gravity, three spacecraft will enter their phase-modulation orbit and eventually achieve a stable 120°-separated uniform distribution in Earth orbit via a phase-modulation maneuver, allowing stereoscopic observation of the Sun and the Heliosphere from the ecliptic plane.

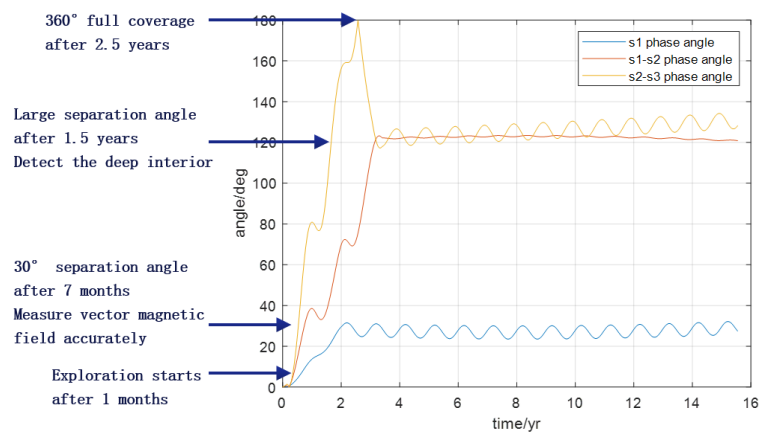

Figure 2. The phase angle between the three spacecraft and Earth.

After the launch, S1 will fly for 839 days, with one lunar flyby to achieve a phase angle of 30° from the Earth, and complete the deployment with a phase holding maneuver at approximately 770m/s; S3 will fly for 1243 days with one lunar flyby to achieve a phase angle of 150° from the Earth, and complete the deployment with a phase holding maneuver at approximately 1220m/s; S2 will fly for 1003.7 days with two lunar flybys to achieve a phase angle of 90° ahead of the Earth, and complete the deployment with a phase holding maneuver at approximately 910m/s. Because the geometry of the three-satellites does not require a strict 120-degree distribution, the number of orbital-keeping maneuvers is limited, about once every six months, lasting a few minutes; the attitude offloading maneuver is scheduled to take place every two days, with a duration of about 10 seconds. All engines startup time is negligible for the mission time and will not interrupt the observations. Details of the orbit include:

(1) Phase angle between the spacecraft and Earth: the phase between S1 and S2 and between S2 and S3 will reach 120 degrees after 3.2 years off the rocket. With orbital maneuvers, the phase angle will remain constant for the following 1.8 years with a 5° fluctuation. About 2.3 years after S1 departs the rocket, the angle between S1 and the Earth will approach 30°, with a varied range of 25-32°. The change in phase angle between S1, S2, S3 and the Earth is shown in Figure 2.

(2) Distance between the spacecraft and the Earth: after separating from the rocket, the distances between the three spacecraft and the Earth increase gradually, and after achieving a 120°-separated uniform distribution, the distances between the three spacecraft and the Earth remain almost constant. The maximum distances between S1, S2, S3 and the Earth are 0.52 AU, 1.41 AU, and 1.82 AU, respectively.

(3) Distance between the spacecraft and the Sun: the distances between the spacecraft and the Sun fluctuate from 0.9 to 1.2 AU before achieving a uniform layout of 120° separation in Earth's orbit, and then they will be 0.97-1.03 AU.

As shown in Figure 2, although it will take 3.2 years to realize the final 120°-separated layout, the payloads will start to carry out scientific missions once they enter the phase-modulation orbit, about one month after launch. About 7 months later, the separation angle between S1 and S2, S2 and S3, and S1 and Earth will reach 30° successively, realizing accurate measurement of vector magnetic field. One and a half years later, the separation angles between the spacecraft become large enough to explore the deep solar interior. Two and a half years after the launch, the full 360° coverage of the Sun can be achieved.

### 4.2 Spacecraft Design

The spacecraft is composed of eight subsystems, including structure and mechanism, thermal control, attitude and orbit control, power supply, general circuit, housekeeping, TT & C communication, and scientific payloads. The configuration of the spacecraft is shown in Figure 3. The spacecraft body adopts a frame-panel configuration. The coordinate system of the spacecraft is defined as follows: Origin of the coordinate (O) is at the center of the spacecraft-rocket separation surface; X axis is parallel to the spacecraft-rocket separation surface, corresponding to the optic axis of the payloads; Y axis is orthogonal to the X and Z axis according to right-handed coordinate system; Z axis is perpendicular to the spacecraft-rocket separation surface, from the rocket to the spacecraft.

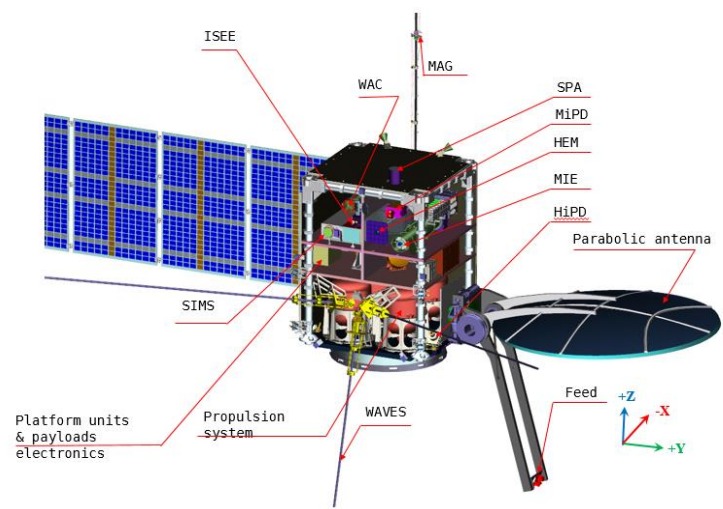

Figure 3. The design of the spacecraft.

In consideration of thermal control and parallelism assurance, all payload units pointing to the Sun are uniformly installed on the +X side of the spacecraft, including SIMS, MIE, HEM,

ISEE, WAC, WAVES, MiPD and HiPD. MAG is placed on the boom on the -X side. The -Z side of the spacecraft is the interface between the spacecraft and the rocket. SPA and the star sensor are installed on it. A solar cell array is installed on the -Y side of the spacecraft, and the solar cells face the +X side after unfolding. A large-diameter antenna with its feed emitting signals at the focal point is installed on the +Y side, and its angle is adjusted and retracted through a one-dimensional rotating mechanism.

According to the launch orbit and weight constraints of the spacecraft, the LM-3B launch vehicle is preliminarily selected to launch the spacecraft in the form of "three spacecraft with one rocket". The requirements of the spacecraft for the launch vehicle system are as follows: (1) The stacked layout of three spacecraft is housed in the effective space of the rocket nose fairing and connected with the supporting cabin of the rocket by a belt and an adapter. The envelope size at launch is Φ3025mm (XOY) × 6500mm (Z), which satisfies the constraints of the fairing of the LM series launch vehicles. (2) The carrying system is capable of launching spacecraft of at least 2690 kg into the Earth-moon transfer orbit at an inclination of 28.5°. (3) At launch, the +Xs axis of the spacecraft points to the head of the rocket(the direction of the rocket movement), while at the time of spacecraft-rocket separation, the -Z axis of the spacecraft points to the Earth satellite-rocket separation. (4) The standard Φ1194 interface will be used for separation. The separation mode is to unlock the package connection, and the reverse rocket or spring separation will be used to support the separation.

The technical specifications of the spacecraft system are summarized in Table A3 in the Appendix.

### 4.3 Data Transmission Rate

According to the data transmission requirements of the aforementioned payloads, and with a transmission time of 4 hours per day, the data transmission rate of S1, S2, and S3 should not be less than 7.5096 Mbps, 1.8378 Mbps, and 1.1154 Mpbs, respectively. To that end, the spacecraft's high gain antenna aperture is designed to be 2.5 M, and a dual power amplifier module with 120W transmitting power is used. The 66M deep-space antenna in Jiamusi receives the majority of the data, with the deep space station of Kashgar, Argentina, and national astronomical observatories providing backup (Figure 4).

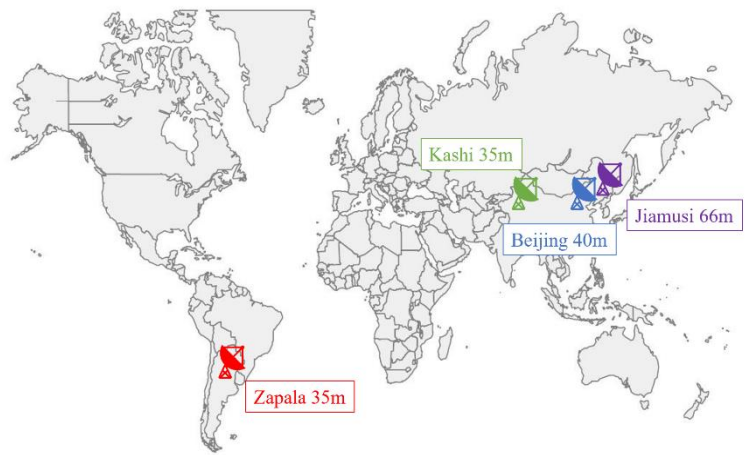

Figure 4. The deep-space network of China can basically achieve full coverage of the data

**4.4 Mission Implementation Scheme**

The overall organization chart for the development of the SOR mission is shown in Figure 5. Now the mission just finished the Phase A study. The overall development of the SOR mission will follow the classical Phase B-C-D approach according to the planning given in Figure 6. The actual start year depends on the funding issue. Once the mission is selected, it would take about 5 years for implementation before the launch.

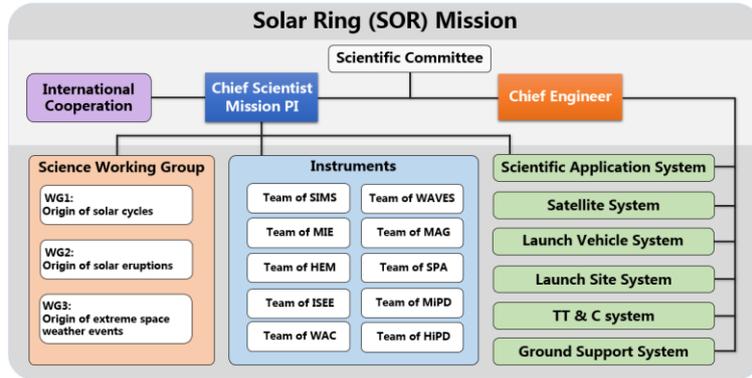

Figure 5. SOR organization chart.

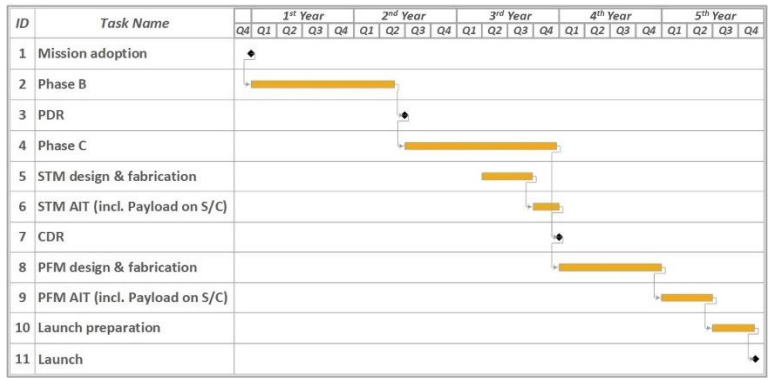

Figure 6. SOR schedule summary.

Besides, the design lifetime of the SOR mission is 5 years after the launch. Based on the existing space missions, e.g., SOHO and STEREO, the actual implementation lifetime is expected to exceed 15 years being able to cover a whole solar cycle to achieve all the scientific goals of the mission.

**5. Summary**

In summary, SOR will deepen our understanding of the Sun and interplanetary space from a variety of perspectives through the following unprecedented capabilities:

(1) Multi-vantage observations of the full-disk magnetic field and velocity field will be realized for the first time. This will help in solving some long-standing problems brought by the line-of-sight projection effect, in understanding the spatial distribution and temporal evolution of the velocity and magnetic field at different scales in the convection zone and the tachocline of the Sun. It will also enable the continuous track of the evolution of active regions to advance in the understanding of the generation and emergence of magnetic fluxes, and boost studies of the dynamics and magnetic field distribution in the solar interior to provide constraints for the establishment of new dynamo theories and models.

(2) Real-time 360° photospheric vector magnetograms will be obtained for the first time. This will provide the most essential input for studying the real-time dynamics of the 3D magnetic structures in the solar atmosphere and establishing data-driven 3D MHD simulations for space weather forecasts. It will meanwhile be able to provide long-term accurate observation of the vector magnetic field and coronal 3D structures in regions of interest, to contribute to establishing new 3D physical models of the origin and evolution of solar eruptions.

(3) Panoramic maps of solar wind disturbances over the low-to-middle latitudes of the inner heliosphere will be available for the first time. Combining with multi-vantage in-situ observations, it will provide unprecedented observational data for studying the formation, evolution, longitudinal distribution and space weather effects of solar energetic phenomena and large-scale disturbances, and therefore enhance the ability of humans to understand and predict extreme space weather events.

Moreover, the innovation of SOR is further manifested by its unique configuration:

(1) The nearest spacecraft, S1, to the Earth is 30°, instead of the conventionally proposed 60°, upstream of the Earth. This would come to a balance between the efficacy and accuracy of space weather forecast. The 30° separation allows a forecast in advance of about 2 days and 6 hours, and compared with spacecraft separated from Earth by a larger angle, the accuracy of the forecast will be less affected by the longitudinal evolution of solar wind as the Sun rotates. Moreover, this will also allow joint observations with spacecraft near the Earth and/or ground-based telescopes, facilitating potentially unprecedented observations.

(2) The spacecraft, S2, 90° downstream of the Earth can not only provide a global picture of the evolution of Earth-directed solar eruptions, but also monitor the space environment for spacecraft on the Hohmann transfer orbit travelling to Mars.

(3) The farthest spacecraft, S3, is 30° away from the Earth-Sun line, which effectively avoids potential communication problems caused by the Sun outage, therefore allowing uninterrupted 360° panoramic observations of the Sun and the inner heliosphere.

With these unprecedented capabilities and the innovative design, we anticipate that the successful implementation of the SOR mission will advance our understanding of the star and the space environment that hold our life, and further foster our capability of expanding the next new territory of human. Any form of international cooperation is welcome.

**Acknowledgement**

This work is supported by the Strategic Priority Program of CAS (Grant Nos. XDB41000000 and XDA15017300) and the Natural Science Foundation of China (NSFC, Grant No. 42188101). Y.W. is particularly grateful for the support of the Tencent Foundation.

Appendix

Table A1. The requirement of payloads defined by the key scientific goals, questions and analyzing methods of SOR

| Scientific Goals | Science Questions | Measurement Methods | SIMS | MIE | HEM | ISEE | WAC | WAVES | MAG | SPA | MiPD | HiPD |
|---|---|---|---|---|---|---|---|---|---|---|---|---|
| Origin of solar cycles | What are the structures and dynamic processes of the solar interior? | Invert the internal structure and dynamical processes based on 360° observations of the full-disk Doppler velocity map | √ | | | | | | | | | |
| | How is surface magnetic flux generated and emerged? | Track magnetic and flow structures at different scales | √ | | | | | | | | | |
| | How is the global magnetic field distributed and evolving? | Perform 360° observations of the full-disk Doppler velocity map and magnetic field | √ | | | | | | | | | |
| Origin of solar eruptions | How are solar eruptions coupled with global coronal and interplanetary magnetic fields? | Reconstruct the 3D structure of the active region and global coronal fields and run data-based models for the eruption process. Compare the model with EUV and white- | √ | √ | √ | √ | | √ | | | √ | √ |
| | How is the eruption energy accumulated | | √ | √ | √ | √ | √ | √ | | | | |

| Scientific Goals | Science Questions | Measurement Methods | SIMS | MIE | HEM | ISEE | WAC | WAVES | MAG | SPA | MiPD | HiPD |
|---|---|---|---|---|---|---|---|---|---|---|---|---|
| | and released? | light observations. | | | | | | | | | | |
| | How are solar energetic particles accelerated? | Analyze the 3D structure and temporal characteristics of the energy-release site. Study the UV, X-ray and radio emissions. | √ | √ | √ | √ | | √ | | | √ | √ |
| Origin of extreme space weather events | How do solar disturbances evolve in the interplanetary space? | Obtain the dynamic solar wind properties based on multi-view images together with radio and in-situ measurements. | √ | | | | √ | √ | √ | √ | √ | √ |
| | How are solar energetic particles transported in interplanetary space? | Obtain 3D structures of CMEs, shocks, CIRs based on multi-view images. Analyze in-situ particle properties and their spatial distribution. | √ | √ | √ | √ | √ | √ | √ | √ | √ | √ |
| | What are the key parameters for space weather forecasts? | Use the upstream data to identify the solar and interplanetary triggers of geomagnetic storms and explore their relationship | √ | √ | √ | √ | √ | √ | √ | √ | √ | √ |

| Scientific Goals | Science Questions | Measurement Methods | SIMS | MIE | HEM | ISEE | WAC | WAVES | MAG | SPA | MiPD | HiPD |
|---|---|---|---|---|---|---|---|---|---|---|---|---|
| | | with geomagnetic storms for the forecast. | | | | | | | | | | |
| | Spacecraft for the corresponding payload | | S1 S2 S3 | S1 | S1 S2 | S1 S3 | S1 S2 S3 | S1 S2 S3 | S1 S2 S3 | S1 S2 S3 | S1 S2 S3 | S1 S2 S3 |

Table A2. Summary of technical specifications of the SOR payloads

| Payload | Spacecraft | | Mass (kg) | Power (W) | | Data Rate (Mbps) | | Observation Requirements | Technical Specifications | Requirements for the spacecraft and constraints | TRL |
|---|---|---|---|---|---|---|---|---|---|---|---|
| | # | Onboard | | Regular | Peek | Regular | Peek | | | | |
| SIMS | S1 | √ | 60 | 50 | 75 | 0.592 | 0.592 | Time resolution for velocity field: 1 min  Pixel Resolution for velocity filed: 4"  Time resolution for vector magnetic field: 15 min  Pixel resolution for vector magnetic field: 1" | Pixel resolution: 1"  Spatial resolution: 2"  Detector pixels: 2048*2048  Observation line: Fe I 532.4 nm  FWHM: 0.011± 0.001 nm  Central wavelength drift: 0.001nm/hour  Wavelength tuning range: [-0.05nm, 0.05nm] | 1) The platform has the ability to point to the sun, with a pointing accuracy of 0.01° (3σ);  2) The platform has pointing stability, and the stability is 0.0005°/s (three axes, 3σ);  3) Thermal interface: the platform provides basic thermal control, 22±5 ℃;  4) Structural interface: the optical box is connected to the satellite thermal insulation through the connecting lugs at the bottom of the box. The thermal insulation pad is made of glass fiber reinforced plastic, the main body is fixed with | 5 |
| | S2 | √ | | 50 | 75 | 0.208 | 0.208 | Time resolution for velocity field: 1 min  Pixel resolution for velocity filed: 4"  Time resolution for vector magnetic field: 60 min | Sensitivity of magnetic field measurements: 10 Gauss for vertical field, 200 Gauss for transverse filed  Time resolution for velocity field: 1 min  Sensitivity of velocity field | | |

| Payload | Spacecraft # | Spacecraft Onboard | Mass (kg) | Power (W) Regular | Power (W) Peek | Data Rate (Mbps) Regular | Data Rate (Mbps) Peek | Observation Requirements | Technical Specifications | Requirements for the spacecraft and constraints | TRL |
|---|---|---|---|---|---|---|---|---|---|---|---|
| | | | | | | | | Pixel resolution for vector magnetic field: 1" | measurements: 30 m/s @ 1 min Pointing stability: pointing to the Sun, 0.5"/15 min | M8 screws, and the heat dissipation plate bracket is fixed with M6 screws; 5) The electric control box is thermally connected to the satellite cabin through the bottom connecting lugs, and the connecting screws are M5 screws; 6) Electrical interface: one-way primary power supply, the power supply voltage is 30±3 V; 7) Telemetry and remote control interface: transmission by CAN bus scheme; 16 channels of analog telemetry acquisition; 8 channels of remote control commands; 8) Data interface: 1 | |
| | S3 | √ | | 50 | 75 | 0.144 | 0.144 | Time resolution for velocity field: 1 min Pixel resolution for velocity filed: 4" Time resolution for vector magnetic field: 15 min Pixel resolution for vector magnetic field: 1" | | | |

| Payload | Spacecraft # | Spacecraft Onboard | Mass (kg) | Power (W) Regular | Power (W) Peek | Data Rate (Mbps) Regular | Data Rate (Mbps) Peek | Observation Requirements | Technical Specifications | Requirements for the spacecraft and constraints | TRL |
|---|---|---|---|---|---|---|---|---|---|---|---|
| | | | | | | | | | | LVDS interface. | |
| MIE | S1 | √ | 40.5 | 35 | 35 | 0.50 | 0.52 | Wavelengths bands: 304/174/130 Å Pixel resolution: 1.5" The whole field of view is 1.6 Rs with a time resolution of about 15 s The region of interest has a field of view of about 0.2 x 0.2 Rs$^2$, with a time resolution of about 2 s | Aperture: 120 mm Focal length: 1 m Camera: 2k×2k back-illuminated CMOS Wavelength bands: 304/174/130 Å Bandwidth: 10-30 Å | 1) The platform has the ability to point to the sun, with a pointing accuracy of 0.01°; 2) The platform has image stabilization capability, and the image stabilization accuracy is 0.0005°/s; 3) Thermal interface: heat insulation installation, CMOS chip heat dissipation, satellite provides scattering surface, the platform provides basic thermal control, 20±5℃; 4) Structural interface: rigidly connected to the satellite platform directly through M6 screws; 5) Electrical interface: | 5 |

| Payload | Spacecraft | | Mass (kg) | Power (W) | | Data Rate (Mbps) | | Observation Requirements | Technical Specifications | Requirements for the spacecraft and constraints | TRL |
|---|---|---|---|---|---|---|---|---|---|---|---|
| | # | Onboard | | Regular | Peek | Regular | Peek | | | | |
| | | | | | | | | | | voltage 30 ± 3 V, current 2 A; preferably 5 V secondary power supply;<br>6) Telemetry and remote control interface: 2 RS422 interfaces;<br>(7) Data interface: 2 LVDS interfaces. | |
| | S2 | × | - | - | - | - | - | - | - | - | |
| | S3 | × | - | - | - | - | - | - | - | - | |
| HEM | S1 | √ | 15 | 18 | 21 | 0.003 | 0.03 | Energy Spectrum & Imaging observation Spectrum energy range: 3 keV-10 MeV Imaging energy range: 3-120 keV Field of view：1.5 Rs | Tungsten grating: 7 pitches covering the resolution range from 8 arcsec to 180 arcsec Collimator length：~85 mm Size(mm*mm*mm)：25×25×900 Detectors: 45 CdTe detectors and 1 | 1) The platform has the ability to point to the sun, and the pointing accuracy is better than 0.01°;<br>2) The platform has image stabilization capability, and the image stabilization accuracy is better than 0.0005°/s;<br>3) Thermal interface: | 6 |

| Payload | Spacecraft | | Mass (kg) | Power (W) | | Data Rate (Mbps) | | Observation Requirements | Technical Specifications | Requirements for the spacecraft and constraints | TRL |
|---|---|---|---|---|---|---|---|---|---|---|---|
| | # | Onboard | | Regular | Peek | Regular | Peek | | | | |
| | | | | | | | | Spatial resolution: ≤8" Energy resolution：1 keV@6 keV Time resolution: regular 5 s, Peek 0.5 s、1 s and 2.5 s | Lanthanum Bromide/Cerium Bromide detector Energy segments: 32- 64 Accuracy of pointing to the Sun: ≤2" | thermally conductive installation, satellite provides scattering surface, the platform provides basic thermal control about 20±5 ℃; 4) Structural interface: tentative lug installation. 5) Electrical interface: voltage 30 V ± 3 V, current 0.7 A; 6) Telemetry and remote control interface: 1 CAN bus interface; 7) Data interface, 1 LVDS interface; 8) The detector has a built-in on-orbit calibration exemption source; 9) Thermal control requirements: the temperature | |
| | S2 | √ | 9 | 15 | 18 | 0.003 | 0.015 | Spectrum & Imaging observation Spectrum energy range: 3 keV-10 MeV Imaging energy range: 3-120 keV Field of view：1.5 Rs Spatial resolution: ≤20" Energy resolution：1 | Tungsten grating: 7 pitches covering the resolution range from 8 arcsec to 180 arcsec Collimator length：~85 mm Size(mm*mm*mm)：25×25×900 Detectors: 39 CdTe detectors and 1 Lanthanum Bromide/Cerium Bromide detector Energy segments: | | |

| Payload | Spacecraft | | Mass (kg) | Power (W) | | Data Rate (Mbps) | | Observation Requirements | Technical Specifications | Requirements for the spacecraft and constraints | TRL |
|---|---|---|---|---|---|---|---|---|---|---|---|
| | # | Onboard | | Regular | Peek | Regular | Peek | | | | |
| | | | | | | | | keV@6 keV Time resolution: regular 5 s, Peek 1 s and 2.5 s | 32- 64 Accuracy of pointing to the Sun: ≤2" | difference between the front and rear mounting substrates is within 2 degrees; the detector temperature is 20±2 ℃. | |
| | S3 | × | - | - | - | - | - | - | - | - | |
| ISEE | S1 | √ | 8 | 25 | 50 | 0.0033 | 0.0033 | Field of View: the whole Sun Time resolution: 1 min Spectrum resolution：≥500 Wavelength range：170-260Å | Field of view: 40' Time resolution: 1min Radiometric Calibration Accuracy: ＜30% Spectrum resolution: ＜0.4 Å/200 Å Signal to noise ratio：>10:1 Pixel spectral resolution: ＜0.1 Å Wavelength range：170-260 Å Detector array size: | 1) The platform has the ability to point to the sun, with a pointing accuracy of 0.01°; 2) The platform has image stabilization capability, and the image stabilization accuracy is 0.0005°/s; 3) Thermal interface: heat insulation installation, CMOS chip heat dissipation, satellite provides scattering surface, the platform provides basic thermal control | 5 |
| | S2 | × | - | - | - | - | - | - | | | |
| | S3 | √ | 8 | 25 | 50 | 0.0011 | 0.0011 | Field of View: the whole Sun | | | |

| Payload | Spacecraft | | Mass (kg) | Power (W) | | Data Rate (Mbps) | | Observation Requirements | Technical Specifications | Requirements for the spacecraft and constraints | TRL |
|---|---|---|---|---|---|---|---|---|---|---|---|
| | # | Onboard | | Regular | Peek | Regular | Peek | | | | |
| | | | | | | | | Time resolution: 3 min<br>Spectrum resolution：⩾500<br>Wavelength range：170-260 Å | 2048x1 | about 20±5 ℃;<br>4) Structural interface: rigidly connected to the satellite platform directly through M6 screws;<br>5) Electrical interface: voltage 30 V ± 3 V, current 2 A; preferably 5 V secondary power supply;<br>6) Telemetry and remote control interface: 2 RS422 interfaces;<br>(7) Data interface: 2 LVDS interfaces. | |
| WAC | S1 | √ | 25 | 80 | 100 | 0.032 | 0.032 | Field of view: 5-45 Rs<br>Wavelength bandwidth: 650-750 nm<br>Spatial resolution: 45"<br>Time | Field of view: 5-45 Rs<br>Wavelength bandwidth: 650-750 nm<br>Spatial resolution: 45"<br>Time resolution: | 1) The platform has the ability to point to the sun, with a pointing accuracy of 0.01°;<br>2) The platform has image stabilization capability, and the image stabilization | 5 |

| Payload | Spacecraft | | Mass (kg) | Power (W) | | Data Rate (Mbps) | | Observation Requirements | Technical Specifications | Requirements for the spacecraft and constraints | TRL |
|---|---|---|---|---|---|---|---|---|---|---|---|
| | # | Onboard | | Regular | Peek | Regular | Peek | | | | |
| | S2 | √ | | 80 | 100 | 0.016 | 0.016 | resolution：<br>-white light 10 min<br>-polarized light 60 min<br><br>Field of view: 5-45 Rs<br>Wavelength bandwidth: 650-750nm<br>Spatial resolution: 45"<br>Time resolution：<br>-white light 20 min<br>-polarized light 120 min | -white light 10 min<br>-polarized light 60 min<br>Exposure time：30 s(white light)；3×90 s(polarized light)<br>Stray light suppression level：$10^{-12}$ B | accuracy is 21 arcsec/30s;<br>3) Installation surface requirements: point to the sun; no obstacle in the field of view ±45 degrees in front of the coronagraph;<br>4) Thermal control: main body temperature of coronagraph: 22 ℃ ±4 ℃ (average temperature of coronagraph frame); storage temperature of the coronagraph in orbit: -10 ℃ ~40 ℃ ; temperature level of CCD chip assembly: -80℃~+22 ℃; thermal installation, CCD chip heat dissipation, satellite provides scattering surface, the | |
| | S3 | √ | | 80 | 100 | | | | | | |

| Payload | Spacecraft | | Mass (kg) | Power (W) | | Data Rate (Mbps) | | Observation Requirements | Technical Specifications | Requirements for the spacecraft and constraints | TRL |
|---------|---|---|---|---|---|---|---|---|---|---|---|
| | # | Onboard | | Regular | Peek | Regular | Peek | | | | |
| | | | | | | | | | | platform provides basic thermal control about 22±5℃; 4) Structural interface: 3 fixed foot structural interfaces; 5) Electrical interface: voltage 30 V ± 3 V, current 2 A; | |
| WAVES | S1 | √ | 16.8 | 24 | 24 | 0.0044 | 0.044 | Observation frequency range 1: 10 kHz~2 MHz, with resolution ≤10 kHz Observation frequency range 2：1-50 MHz, with resolution ≤100 kHz Detection range: 1 – 200 Rs Time resolution: -regular 10 s | Observation frequency range 1: 10 kHz~2 MHz, with resolution ≤10 kHz Observation frequency range 2：1-50 MHz, with resolution ≤100 kHz Dynamic range: ≥70 dB Time resolution：1, 10, 30 s | 1) The platform has the ability to point to the sun, with a pointing accuracy of 0.1°; 2) Thermal interface: the working temperature of the electronic box in the cabin: -20℃～+55℃; the working temperature of the preamplifier outside the cabin: -55 ℃ ～ +80 ℃ ; the working temperature of the | 6 |

| Payload | Spacecraft # | Spacecraft Onboard | Mass (kg) | Power (W) Regular | Power (W) Peek | Data Rate (Mbps) Regular | Data Rate (Mbps) Peek | Observation Requirements | Technical Specifications | Requirements for the spacecraft and constraints | TRL |
|---|---|---|---|---|---|---|---|---|---|---|---|
| | | | | | | | | -peek 1 s | | antenna outside the cabin: -100 ℃ ～ +100 ℃; Antenna pyrotechnics working temperature: -40℃～ +80℃; 3) The antenna is composed of receiving antennas A, B, and C. The antenna rods of the antennas A, B, and C are required to be perpendicular to each other, and the angle between each antenna and the installation surface is the same; 4) Electrical interface: voltage 30 V±3 V, current 1 A; the satellite platform provides 3 circuits of pyrotechnic busbar voltage used by the antenna release | |
| | S2 | √ | | | | 0.0015 | 0.044 | Observation frequency range 1: 10 kHz~2 MHz, with resolution ≤10 kHz Observation frequency range 2：1-50 MHz, with resolution ≤100 kHz Detection range: 1 – 200 Rs Time resolution: -regular 30 s -peek 1 s | | | |
| | S3 | √ | | | | 0.0015 | | Observation frequency range 1: 10 kHz~2 MHz, with resolution ≤10 | | | |

| Payload | Spacecraft # | Spacecraft Onboard | Mass (kg) | Power (W) Regular | Power (W) Peek | Data Rate (Mbps) Regular | Data Rate (Mbps) Peek | Observation Requirements | Technical Specifications | Requirements for the spacecraft and constraints | TRL |
|---|---|---|---|---|---|---|---|---|---|---|---|
| | | | | | | | | kHz Observation frequency range 2：1-50 MHz, with resolution ≤100 kHz Detection range: 1 – 200 Rs Time resolution: 30 s | | structure, voltage 25.5 V±2.5 V, instantaneous current 5~10 A; (5) Telemetry and remote control interface: 1 CAN bus interface; (6) Data interface: 2 LVDS interfaces. | |
| MAG | S1 | √ | 2.5 | 5 | 5 | 0.0025 | 0.0095 | Dynamic range: ±2000 nT Noise: ≤0.01 nT/√Hz@1Hz Sampling rate：-regular 1, 32 Hz –peek 128 Hz | Dynamic range: ±2000 nT Resolution: ≤0.01 nT Noise: ≤0.01 nT/√Hz@1Hz Sampling rate: 1 32 and 128 Hz | 1) The platform provides an extension rod mechanism (deployed length ≥ 6m), one magnetometer sensor is installed on the top of the extension rod, and the other sensor is installed on the extension rod about 1m away from the top. After the extension rod | 7 |
| | S2 | √ | | | | 0.0025 | 0.0025 | Dynamic range: ±2000 nT Noise: ≤0.01 nT/√Hz@1Hz Sampling rate： | | | |
| | S3 | √ | | | | | | | | | |

| Payload | Spacecraft | | Mass (kg) | Power (W) | | Data Rate (Mbps) | | Observation Requirements | Technical Specifications | Requirements for the spacecraft and constraints | TRL |
|---|---|---|---|---|---|---|---|---|---|---|---|
| | # | Onboard | | Regular | Peek | Regular | Peek | | | | |
| | | | | | | | | 1, 32 Hz | | mechanism is deployed, the sensor can be kept away from the satellite body, and away from thrusters, plumes, wakes, solar panels, etc. The electronics box is installed in the cabin nearby; 2) Thermal interface: the electronic box is thermally installed, and the operating temperature of the installation surface is -40~+45 ℃; 3) Structural interface: the mounting surface of the electronics box is about 220 mm×170 mm, and the two sensors are installed on the extension rod mechanism adiabatically; | |

| Payload | Spacecraft | | Mass (kg) | Power (W) | | Data Rate (Mbps) | | Observation Requirements | Technical Specifications | Requirements for the spacecraft and constraints | TRL |
|---|---|---|---|---|---|---|---|---|---|---|---|
| | # | Onboard | | Regular | Peek | Regular | Peek | | | | |
| | | | | | | | | | | 4) Electrical interface: voltage 30 V±3 V, power consumption ≤5 W; (5) Telemetry and remote control interface: 1 RS422 interface; (6) Data interface: 1 RS422 interface. | |
| SPA | S1 | √ | 6 | 20 | 20 | 0.0022 | 0.012 | Energy range: 100 eV - 30 keV Mass resolution: $H^+$, $He^+$ Time resolution: ≤32s Energy resolution: ≤15% | Energy range: 100 eV – 30 keV Energy resolution(ΔE/E): ≤15% Mass resolution(Δm/m): ≤20% Field of view: ±30°×±90° Angular resolution: 6°×6° Time resolution: ≤32s | 1) There is no obstruction within the field of view, and the field of view covers the sun; keep away from thrusters, plumes, wakes, etc.; there is no strong electromagnetic field interferences nearby; 2) Thermal interface: thermally conductive installation, the operating temperature of the installation | 7 |
| | S2 | √ | | | | | | | | | |
| | S3 | √ | | | | | | | | | |

| Payload | Spacecraft | | Mass (kg) | Power (W) | | Data Rate (Mbps) | | Observation Requirements | Technical Specifications | Requirements for the spacecraft and constraints | TRL |
|---|---|---|---|---|---|---|---|---|---|---|---|
| | # | Onboard | | Regular | Peek | Regular | Peek | | | | |
| | | | | | | | | | | surface is -40~+45℃; 3) Structural interface: It is installed in the satellite cabin through the opening of the spacecraft skin, and the sensor protrudes 150 mm from the outer surface of the spacecraft skin; 4) Electrical interface: voltage 30 V±3 V, power consumption ≤20 W; 5) Telemetry and remote control interface: 1 RS422 interface; 6) Data interface: 1 RS422 interface | |

| Payload | Spacecraft # | Spacecraft Onboard | Mass (kg) | Power (W) Regular | Power (W) Peek | Data Rate (Mbps) Regular | Data Rate (Mbps) Peek | Observation Requirements | Technical Specifications | Requirements for the spacecraft and constraints | TRL |
|---|---|---|---|---|---|---|---|---|---|---|---|
| MiPD | S1 | √ | 3.5 | 4 | 4 | 0.0002 | 0.0008 | Time resolution: -regular 1 min -peek 10 s Energy range for electrons: 40-600 keV Energy channels: 128 | Energy range for electrons: 40-600 keV Field of view @ 3 directions: 3×30°×40° Sampling rate: 1s | 1) There is no obstruction within the field of view; keep away from thrusters, plumes, wakes, etc. Avoid direct sun exposure to openings in the probe collimator. Keep away from equipment with permanent magnets; 2) Thermal interface: thermally conductive installation, the working temperature of the installation surface is -40~+45 ℃, the best working temperature: -10~+25 ℃; 3) Structural interface: It is installed in the satellite cabin through the opening of the spacecraft skin, and | 7 |
| | S2 | √ | | | | | | | | | |
| | S3 | √ | | | | | | | | | |

| Payload | Spacecraft # | Spacecraft Onboard | Mass (kg) | Power (W) Regular | Power (W) Peek | Data Rate (Mbps) Regular | Data Rate (Mbps) Peek | Observation Requirements | Technical Specifications | Requirements for the spacecraft and constraints | TRL |
|---|---|---|---|---|---|---|---|---|---|---|---|
| | | | | | | | | | | the sensor protrudes from the outer surface of the spacecraft skin by about 100~130 mm;<br>4) Electrical interface: voltage 30 V±3 V, power consumption ≤4 W;<br>5) Telemetry and remote control interface: 1 RS422 interface;<br>6) Data interface: 1 RS422 interface. | |
| HiPD | S1 | √ | 3.5 | 5 | 5 | 0.003 | 0.008 | Energy range for protons: 50 keV – 300 MeV Energy range for αparticles and heavy ions: | Energy range for protons: 50 keV – 300 MeV Energy range for αparticles and heavy ions: 10-100 | 1) There is no obstruction within the field of view; keep away from the thruster, plume, wake, etc.;<br>2) Thermal interface: | 8 |

| Payload | Spacecraft | | Mass (kg) | Power (W) | | Data Rate (Mbps) | | Observation Requirements | Technical Specifications | Requirements for the spacecraft and constraints | TRL |
|---|---|---|---|---|---|---|---|---|---|---|---|
| | # | Onboard | | Regular | Peek | Regular | Peek | | | | |
| | S2 | √ | | | | | | 10-100 MeV/n Energy range for electrons: 100 keV – 12 MeV Energy channels: 128 Directions of measurements: 2 Time resolution: -regular 1 min -Peek 10 s | MeV/n Energy range for electrons: 100 keV – 12 MeV LET: 0.1 - 37 MeV/(mg/cm$^2$) Directions of measurements: 2, towards and away from the sun Angular of FOV：>45° Sampling rate: 1 s | thermally conductive installation, the operating temperature of the installation surface is -40~+45 ℃; 3)Structural interface: It is installed in the satellite cabin through the opening of the spacecraft skin, and the sensor protrudes from the outer surface of the spacecraft skin by about 20 mm; 4) Electrical interface: voltage 30 V±3 V, power consumption ≤5 W; 5) Telemetry and remote control interface: 1 RS422 interface; 6) Data interface: 1 RS422 interface. | |
| | S3 | √ | | | | | | | | | |
| Total | S1 | Summar | 180.8 | 266 | 339 | 1.1447 | 1.2516 | | | 1) The observation | |

| Payload | Spacecraft | | Mass (kg) | Power (W) | | Data Rate (Mbps) | | Observation Requirements | Technical Specifications | Requirements for the spacecraft and constraints | TRL |
|---|---|---|---|---|---|---|---|---|---|---|---|
| | # | Onboard | | Regular | Peek | Regular | Peek | | | | |
| | | y | | | | | | | | time synchronization accuracy of the first 4 payloads on the same satellite is better than 0.01 s; 2) The observation time synchronization accuracy of the first six payloads on different satellites is better than 0.01 s. | |
| | | Constraints | 180.8 | 266 | | 1.6667 | | | | | |
| | S2 | Summary | 126.3 | 203 | 251 | 0.2364 | 0.3063 | | | | |
| | | Constraints | 126.3 | 203 | | 0.3200 | | | | | |
| | S3 | Summary | 125.3 | 213 | 283 | 0.1705 | 0.1859 | | | | |
| | | Constraints | 125.3 | 213 | | 0.1930 | | | | | |

Table A3. The major technical specifications of the spacecraft

| Categories | Index term | Target requirement |
|---|---|---|
| Orbit | Launch trajectory | Earth-moon transfer orbit, Inclination: 28.5° |
| | Working orbit | Three evenly distributed spacecraft with a separation of 120° in the Earth's orbit. S1: 30° upstream of the Earth, S2: 90° downstream of the Earth, S3: 150° upstream of the Earth. |
| Power consumption | Platform Power consumption | S1: 932W, S2: 863W, S3:874W |
| | Supported payload power consumption | S1: 266W, S2: 203W, S3: 213W |
| Weight | Platform weight | S1: 870 kg, S2: 845 kg, S3: 975 kg |
| | Supported payload weight | S1:180.8 kg, S2:126.3 kg, S3:125.3 kg |
| Size | at launch | Single spacecraft: Φ3025 mm×2239 mm  Three spacecraft with one rocket: Φ3025 mm×6500 mm |
| | in orbit | Single spacecraft: 9749 mm(X)×7540 mm(Y)×6560 mm(Z) |
| Attitude control | Pointing pattern | Pointing to the sun, three-axis stability |
| | Pointing accuracy | Better than 0. 01 degrees |
| | Pointing stability | Better than 0.0005°/s (three axes, 3σ) |
| Orbit control | maneuver capability | S1:0.77 km/s, S2:0.91 km/s, S3: 1.22 km/s |
| | Orbit accuracy | Better than 5‰ |
| Thermal control | Platform temperature | About -10 ° ~ + 45 ° |
| | Temperature in the payload bay | 20 ± 5 ° |
| Structure | Strength and stiffness | Satisfy the condition of LM-3B |
| | Interface with the carrier | Standard 1194 packet-tape interface |
| | Installation error of payload axis | 1' |
| Energy | Bus Voltage | 28 ± 1V |
| | Lithium-ion battery | 30Ah, 5 years |
| Measurement and control | Measurement and control system | X-band measurement and control system |
| | Uplink rate | 2000 bps |
| | Downlink rate | 16384 bps |
| Spacecraft housekeeping system | Processing speed | Processing speed should not be lower than 15 MIPS, and support floating-point operation |
| Data transmission | Frequency Band | X-band |
| | Storage capacity | S1: 150 Gbits, S2/S3: 30 Gbits |
| | Data transmission | S1: 10, S2: 1.92, S3:1.158(Mbps) @Jiamusi/Wuqing |